\documentclass[epj]{webofc}
\usepackage[varg]{txfonts} 
\usepackage{graphicx,wrapfig,lipsum}
\usepackage{amssymb,amsmath}
\usepackage{subeqnarray}
\usepackage{hyperref}
\newcommand{\greeksym}[1]{{\usefont{U}{psy}{m}{n}#1}}

\newcommand{\udelta}{\mbox{\greeksym{d}}}

\newcommand{\uLambda}{\mbox{\greeksym{L}}}

\newcommand{\meV}{\ensuremath{\,{\rm meV}}}

\newcommand{\MeV}{\ensuremath{\,{\rm MeV}}}

\newcommand{\Neff}{\ensuremath{{N_{\rm eff}}}}

\newcommand{\nc}{\newcommand} 
\nc{\req}[1]{Eq.\,(\ref{#1})} \nc{\reqp}[1]{Eq.\,(\ref{#1}) on page \pageref{#1}}  
\nc{\rf}[1]{Fig.~\ref{#1}} \nc{\rfp}[1]{Fig.~\ref{#1} on page \pageref{#1}}  
\nc{\rt}[1]{table~\ref{#1}} \nc{\rtp}[1]{table~\ref{#1} on page \pageref{#1}}  
\nc{\Th}{\ensuremath{T_\mathrm{H}\,}}
\nc{\pp}{\ensuremath{pp\ }}
\nc{\pA}{\ensuremath{p A\ }}
\nc{\hAA}{\ensuremath{AA\ }}
\nc{\D}{\mathrm{d}}
\nc{\E}{\mathrm{e}}

\def\beq{\begin{equation}}
\def\eeq{\end{equation}}
\newcommand{\beqn}{\begin{equation}}
\newcommand{\eeqn}{\end{equation}}

\wocname{EPJ Web of Conferences}
\woctitle{ICFNP2015}
\begin{document}
\title{The hot Hagedorn Universe}
\subtitle{
Presented at the ICFNP2015 meeting, August 2015}

\author{Johann Rafelski\inst{1}%
\and
  Jeremiah Birrell\inst{2}%
}

\institute{Department of Physics, The University of Arizona, Tucson, AZ 85721, USA
\and
   Department of Mathematics, The University of Arizona, Tucson, AZ 85721, USA
  }

\abstract{%
In the context of the half-centenary of Hagedorn temperature and the statistical bootstrap model (SBM) we present a short account of how these insights coincided with the establishment of the hot big-bang model (BBM) and helped resolve some of the early philosophical difficulties. We then turn attention to the present day context and show the dominance of strong interaction quark and gluon degrees of freedom in the early stage, helping to characterize the properties of the hot Universe. We focus attention on the current experimental insights about cosmic microwave background (CMB) temperature fluctuation, and develop a much improved understanding of the neutrino freeze-out, in this way paving the path to the opening of a direct connection of quark-gluon plasma (QGP) physics in the early Universe with the QCD-lattice, and the study of the properties of QGP formed in the laboratory.
}
\maketitle
\section{The big-bang model established}
\label{EarlyHagedorn}
Who today can remember that before 1965 the big-bang model (BBM) was challenged by those who had difficulty accepting that there is a beginning of time? And, even after the discovery of the cosmic microwave background (CMB) radiation, announced May 1965, the idea that the primordial Universe had to be hot generated an additional challenge: how could all the energy in the Universe come from an initial space-time singularity? Such conceptual difficulties were exploited by those who did not like the hot BBM model. One of us can remember the scientific disputes. 

A literature search shows that Hagedorn\rq s statistical bootstrap model (SBM) was an inadvertent solution to many conceptual difficulties. In fact, a who\rq s who   of cosmology of this period cites Hagedorn\rq s work~\cite{Hagedorn:1965st}. While it could sound presumptuous to claim that Hagedorn was the one who turned the corner in establishing the hot BBM as the standard cosmological model, at the least his contribution was very important. 

The Hagedorn SBM had a built-in feature that was needed, a divergence in the energy content even for a initial singular point volume. The way this works is rather easy to explain. It is convenient to introduce the hadron mass spectrum $\rho(m)$, where
\begin{equation}\label{2chap2eq3.3}
\rho(m)\D m =\mbox{number of hadron states in $\{m,m\!+\!\D m\}$}\;.
\end{equation}
In Hagedorn\rq s SBM, an exponentially growing mass spectrum $\rho(m)\propto m^{-a}\exp (m/\Th)$, is a natural solution. Thus we have within the partition function of a hadron gas comprising many different particles as described by the mass spectrum, a singular contribution
\begin{equation}\label{2chap2eq3.4}
\Delta\, Z(T,V) = \exp\left[V\left(\frac{T}{2\pi}\right)^{3/2} \int_{3\Th}^\infty \rho(m)m^{3/2}\E^{-m/T}\D m\right]\;.
\end{equation} 
The lower limit 3\Th\ on the integral $dm$ ensures that the non-relativistic limit shown in \req{2chap2eq3.4} is reasonable. We can compute, for a few selected values of the power index $a$ in $\rho(m)$ of immediate physical interest, the singular properties of the hadron gas near to $T=\Th+\Delta T$ to be as shown in the table~\ref{gasHG}.

\begin{table}
\centering
\caption{Thermodynamic quantities assuming exponential form of hadron mass spectrum with pre-exponential index $a$: pressure $P$, energy density $\varepsilon$, energy density fluctuation $\udelta\varepsilon/\varepsilon$, the heat capacity $C_V=\D\varepsilon/\D T$ and the sound velocity $v^2=\D P/\D\varepsilon$}
\label{gasHG} 
\begin{tabular}{lccccc}
\hline
\phantom{\Large$\frac{1}{1}$\hspace{-4pt}}$a$ & $P$ & $\varepsilon$ & $\udelta\varepsilon/\varepsilon$
 & $C_V=\D\varepsilon/\D T$ & $v^2=\D P/\D\varepsilon$ \\
\hline\hline
\phantom{\Large$\frac{1}{1}$\hspace{-5pt}}5/2 & $C\ln(T_0/\varDelta T)$ & $C/\varDelta T$ & $C$ & $C/\varDelta T^2$ & $C\varDelta T$
\\\hline
\phantom{\Large$\frac{1}{1}$\hspace{-5pt}}3 & $P_0-C\varDelta T^{1/2}$ & $C/\varDelta T^{1/2}$ & $C/\varDelta T^{1/4}$ & $C/\varDelta T^{3/2}$ & $C\varDelta T$
\\\hline
\phantom{\Large$\frac{1}{1}$\hspace{-5pt}}7/2 & $P_0-C\varDelta T$ & $\varepsilon_0$ & $C/\varDelta T^{1/2}$ & $C/\varDelta T$ & $C\varDelta T$
\\\hline
\phantom{\Large$\frac{1}{1}$\hspace{-5pt}}4 & $P_0-C\varDelta T^{3/2}$ & $\varepsilon_0-C\varDelta T^{1/2}$ & $C/\varDelta T^{3/4}$ & $C/\varDelta T^{1/2}$ & $C\varDelta T$\\
\hline
\end{tabular} 
\end{table}

The value $a=5/2$ which Hagedorn advocated in late 1960s shows a singular energy density and a slower logarithmically divergent pressure. This allows the Universe to originate in a singular volume condition, and yet to contain an infinite energy. This was a new type of singularity, and a revolutionary insight that impacted a few other areas of physics. We see the appearance of the term \lq\lq Hagedorn temperature\rq\rq\ in distant physics topics today. 

Hagedorn was accordingly presenting his ideas also at cosmology meetings and one of the best accounts of his work on hot hadronic matter and the BBM are his lectures at the Cargese Summer school in 1971~\cite{Hagedorn:1973rva}, a classic of the period with very important contributions by other cosmology and astrophysics luminaries.
 
To close this introduction let us quote from one of Hagedorn\rq s popular lectures of 1968: one of last paragraphs reads~\cite{Hagedorn:2016nrh}: \lq\lq At least a few theories about the beginning of the Universe assume a BBM, that is to say a creation explosion. Following previous ideas -- based on traditional black body radiation -- the Universe began with infinite energy density, with energy density proportional to the pressure, and infinitely high temperature. Under such extreme conditions, traditional black body radiation no longer remains, but rather the conditions are found akin to the high-energy collisions of nucleons. And then when strongly interacting matter is present the temperature cannot be infinite, but only about $10^{12}$\,K, and the pressure is not anymore proportional to the energy density but only proportional to its logarithm. This is a different scenario of the beginning of the Universe than was previously thought.\rq\rq\ 

The following is the last paragraph, equally interesting, a typical \lq Hagedornian\rq, and perhaps should be saved for the end of this lecture, but it seems that it should better be presented together with the former content: \lq\lq I (Hagedorn) close with an anecdote: On the bulletin board of a German university the following could once be read among lecture announcements: {\it Tuesdays 9-11 AM, free for all discussion session about the structure of the Universe -- only for the advanced. signed X. } We will, alas, always be beginners.\rq\rq\

\section{The particle and nuclear Universe}
\label{QGPHagedorn}
In the following three sub-sections we will explore key features of the early Universe from the present day perspective, connecting the field of hot hadron matter and quark gluon plasma (QGP) to the observation of the early Universe by way of the CMB. We embed our study into the Universe evolution period during the time window where the laws of nuclear science govern the result. We summarize the pivotal technical details that one can call the standard model of cosmology in the appendices: we describe the cosmology in context of the Robertson-Walker Universe, see Appendix~\ref{scosmo}; derive key relations that characterize the Universe, specifically the thermal and free-streaming components of matter and radiation, in Appendices~\ref{mcosmo}, and~\ref{fcosmo}, respectively. All these inputs allow us to evaluate ab-initio the dynamics of the Universe. This report updates and extends earlier related efforts~\cite{Fromerth:2002wb,Fromerth:2012fe,Rafelski:2013obw,Rafelski:2013yka}.
\subsection{Hot Hagedorn Universe}
Forwarding from the past to the present, we look again at the entries in table~\ref{gasHG}: we note that it is not necessary that the energy density is singular at $T=\Th$; in fact should $a\ge 7/2$, we clearly see that the hadron gas singularity is dissolved. A fit to the hadron mass spectrum of the exponential shape predicts $\Th(a=7/2)=151$\,MeV and $\Th(a=8/2)=144$\,MeV. This value is also confirmed by lattice gauge numerical evaluation; for details of this complicated situation we refer to the discussion in the review, Ref.\,\cite{Rafelski:2015cxa}. 

It is an undisputed fact emerging from the lattice gauge numerical evaluation of strong interaction properties that a singularity at \Th\ is avoided. We find a smooth transformation, a position clearly taken by Hagedorn as of the early 1980s, see for example Ref.\,\cite{Hagedorn:1984hz}.
We can say that hadrons dissolve into their constituents, quarks, and that this phase transformation has been established near $\Th=145+-7$\,MeV~\cite{Borsanyi:2012rr,Borsanyi:2013bia}. An interesting feature at the condition is that the sound velocity vanishes, first discovered in the SBM model, see last column in table~\ref{gasHG}. This has also been confirmed by lattice gauge theory. For this reason the QGP decay into hadrons is subject to non-trivial slow flow dynamics; one can argue that the fireball of hot matter sits and spits out its content that streams out freely. This feature goes well with the sudden hadronization model, but this is not our present discussion topic.

\begin{figure}
\centering
\includegraphics[width=\columnwidth,clip]{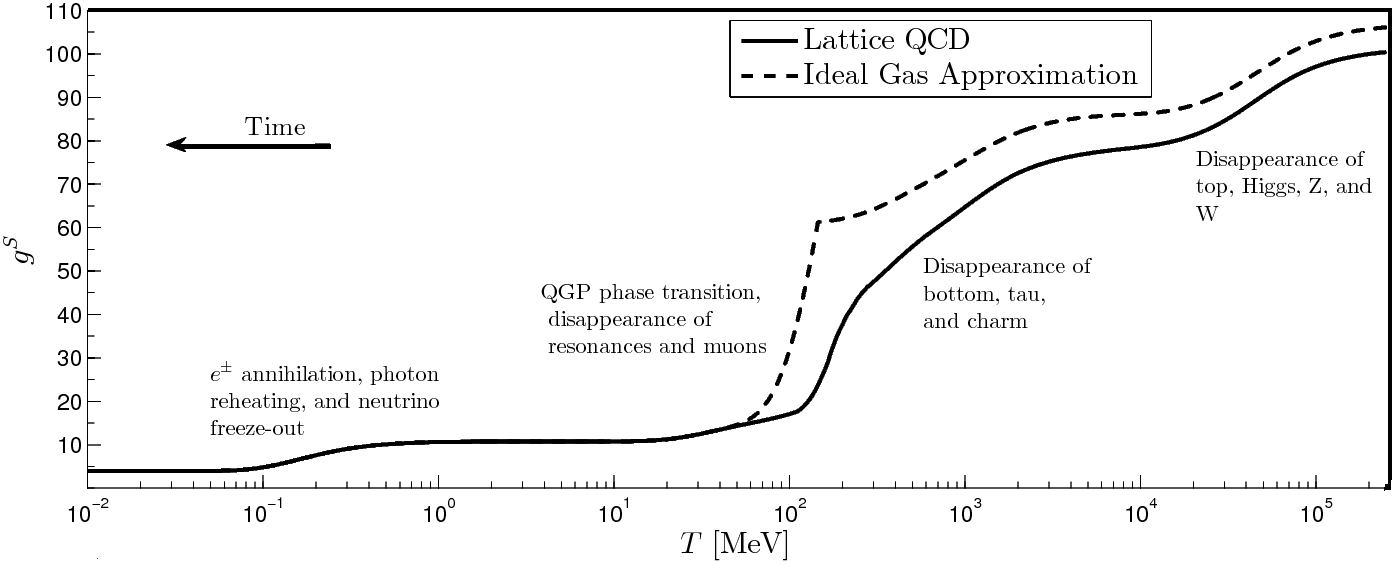}
\caption{Active entropic degrees of freedom, $g^S$, in the Universe as a function of ambient temperature, $T$, from the electroweak transition on right, to conditions similar to present day, on left (time arrow from right to left). Note a significant modifications of the ideal gas (dashed line) model in the QGP era. The extrapolation for $T>500\MeV$ relies on thermal-QCD approach~\cite{MStricklandPrivate}.
\label{fig-geffecS} } 
\end{figure}

We are interested in connecting the properties of the QGP stage of the early Universe to the present day world we see around us today. To achieve this objective, we have to evolve the Universe dynamically. In figure~\ref{fig-geffecS} we show, as an example of such a computation, the evolution of the degrees of freedom as a function of temperature. By far the fastest and most dramatic change occurs when the two phases of strongly interacting matter meet: the Hagedorn gas phase, and the deconfined QGP, with a cross-over point at $T=\Th$. To describe this situation in this computation we use the Wuppertal-Budapest lattice QCD equations of state~\cite{Borsanyi:2012rr,Borsanyi:2013bia} continued to yet higher $T>500$\,MeV based on thermal-QCD~\cite{MStricklandPrivate}. The strong interactions phases are supplemented by ideal gases of weakly and electromagnetically interacting particles, either coupled or free-streaming (such as cold neutrinos), and we incorporate in this study all the usual elements of the ${\uLambda}$CDM model introduced in Appendix~\ref{cosmo}.

\begin{figure}[t]
\centering
\sidecaption
\includegraphics[width=0.7\columnwidth,clip]{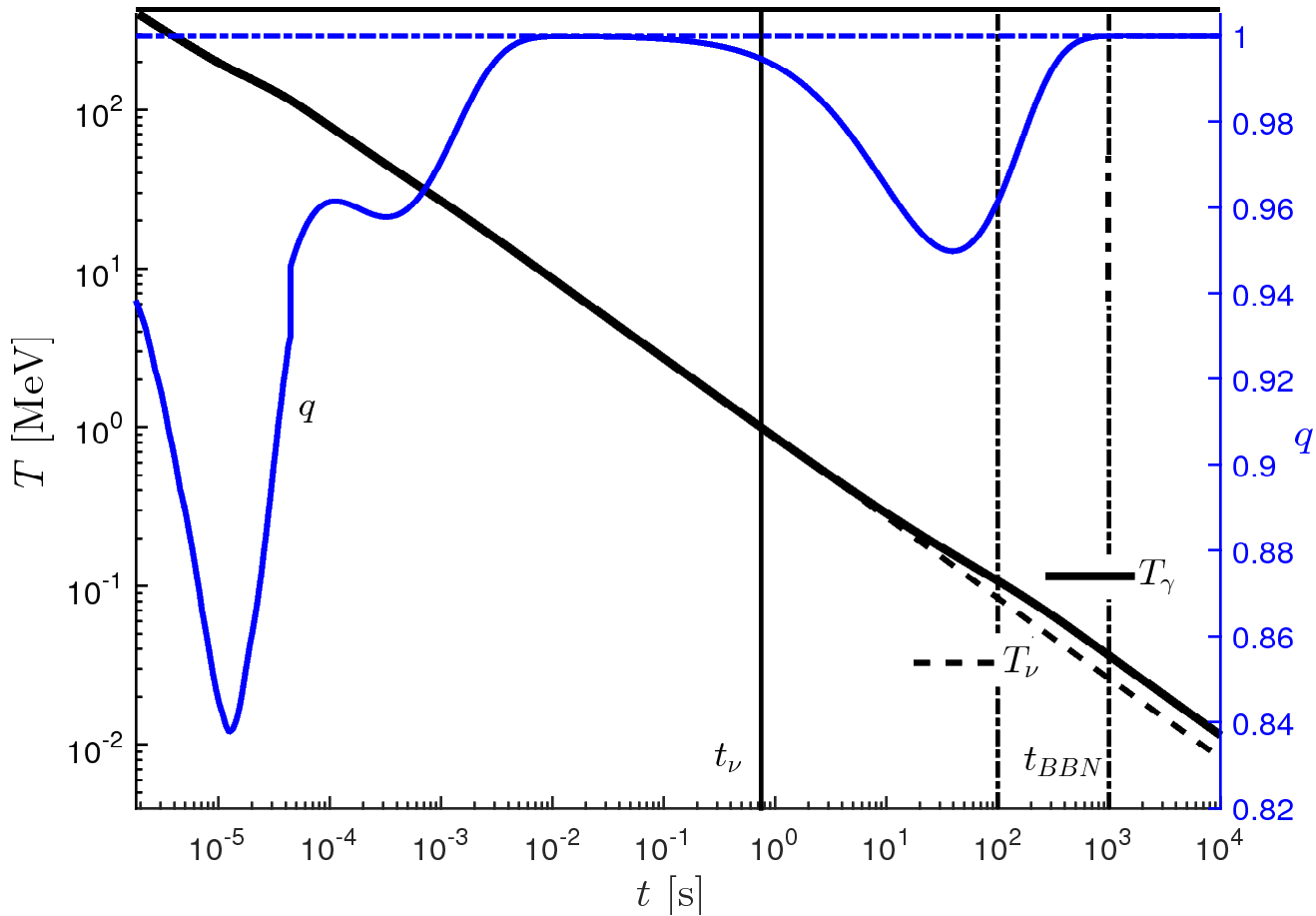}
\caption{The temperature, left scale and deceleration parameter, right scale, as a function of the age of the Universe from 1 micro second to 10,000 seconds (10 orders of magnitude). This is the domain  in the expansion of the Universe where nuclear science plays the primary role.}
\label{fig-qTnuclear}  
\end{figure}

In figure~\ref{fig-qTnuclear} we show the evolution (usual time arrow from left to right) of the Universe temperature as a function of time (left side scale) from midst of QGP at about 400 MeV down to temperature of 10 keV where the BBM nucleosynthesis (BBN) had ended. Note that there are two temperatures shown for large times: the annihilation of electrons with positrons feeds entropy only into the photon degrees of freedom and thus already free-streaming neutrinos are in comparison (dashed line) colder. The sharing of entropy from disappearing (massive) particle families results in reheating of only the coupled degrees of freedom. For reheating to differentiate the temperature, a particle must already have decoupled when annihilation of a massive degree of freedom occurs. It is a phenomenon that therefore does not occur, other than for neutrinos, for other visible (as opposed to dark) particles in the evolution of the Universe. However, dark particle components of dark matter will clearly be much colder and we return to discuss this situation later.

We also show the deceleration parameter $q$ (right side scale) in figure~\ref{fig-qTnuclear}. We saw earlier that $q=1$ corresponds to a radiation dominated Universe where $P/\varepsilon=1/3$. $q$ decreases from unity when the Universe expansion slows down due to emergence of a massive degree of freedom. We also saw that $q=1/2$ for a matter dominated Universe, where pressure $P\ll \varepsilon$. Note the dip around 20 $\mu$s, that is near to hadronization of QGP at \Th, where many very massive baryons and heavy mesons are formed and quickly disappear. Another dip appears due to the pion and muon masses becoming visible, before their particle density also disappears as the Universe cools. A clean version of the same effect is seen for $1<t<300$\,s when in a wide temperature range at the scale of electron mass we see the effect of the electron and positron density. What we learn is that the deceleration parameter $q$ is a diagnostic tool for when a massive particle component in the Universe becomes partially nonrelativistic, but is in an abundance that impacts the speed of expansion.

In the analysis of the CMB, which is the present day experimental access to early the Universe, the particle content in the Universe can be studied, as the speed of expansion impacts the build-up of fluctuation structure. In this analysis one expresses the uncertainty about particle content as an uncertainty in the number of invisible neutrino degrees of freedom $N_\mathrm{eff}$. If there is nothing unusual governing the Universe expansion, within the precision of the data the value $N_\mathrm{eff}\simeq 3$ will be determined. 

During recent years the CMB data analysis yielded $3\lesssim N_\mathrm{eff} \lesssim 4$ with the long term bias above 3.5 broken when the Planck collaboration~\cite{Ade:2013zuv} revised the fit strategy to allow a greater tension with priors, redistributing the error over other parameters of the standard model of cosmology. This is not a suitable place to discuss in any depth the question if there is, or is not, $\delta N_\mathrm{eff}\equiv N_\mathrm{eff}-3>0$. Only the next generation of experiments that allow a significantly greater precision can answer this. Awaiting this and in consideration of the bias towards $\delta N_\mathrm{eff}\simeq 0.3$-0.6, we believe that the period of Universe expansion shown in figure~\ref{fig-qTnuclear} should be reexamined from the nuclear science perspective in order to understand how different nuclear science effects could have an impact on $\delta N_\mathrm{eff}$. 

In the following we address  two sources of $\delta N_\mathrm{eff}$ excess. The first originates in the  generation of unknown light particles during QGP hadronization, and the other due to modification of neutrino freeze-out dynamics. Both effects can only be present if some new physics comes along. We discuss these two effects in turn, beginning with discussion of how the QGP condition can impact the interpretation of CMB temperature fluctuations.

\subsection{New particles from QGP hadronization}\label{QGPsterile}
The fact that $0<\delta N_\mathrm{eff}<1$ does not mean that there could not be one, or more, particle degrees of freedom that are the cause. As long as the particle concerned decouples within QGP we can be sure of a considerable reheating effect, given the excess QGP entropy flow from disappearing degrees of freedom into the remaining visible Universe content. This effect dilutes the free-streaming contribution of the here called \lq secret\rq\ particle considered. While we invented this effect independently and study it in a model-independent way~\cite{Birrell:2014cja}, two earlier and independent efforts can be found in literature, one focused on sterile neutrinos~\cite{Anchordoqui:2011nh,Anchordoqui:2012qu}, and the other on broken dark matter symmetries~\cite{Weinberg:2013kea}. All this work assumes the presence of secret particles emerging from QGP that remain undetected today. That it must be happening somewhere in QGP is obvious considering the systematic behavior seen in figure~\ref{fig-geffecS} as the number of degrees of freedom drops rapidly when temperature drops towards 100 MeV. 

\begin{figure}
\centering
\sidecaption
\includegraphics[width=0.65\columnwidth,clip]{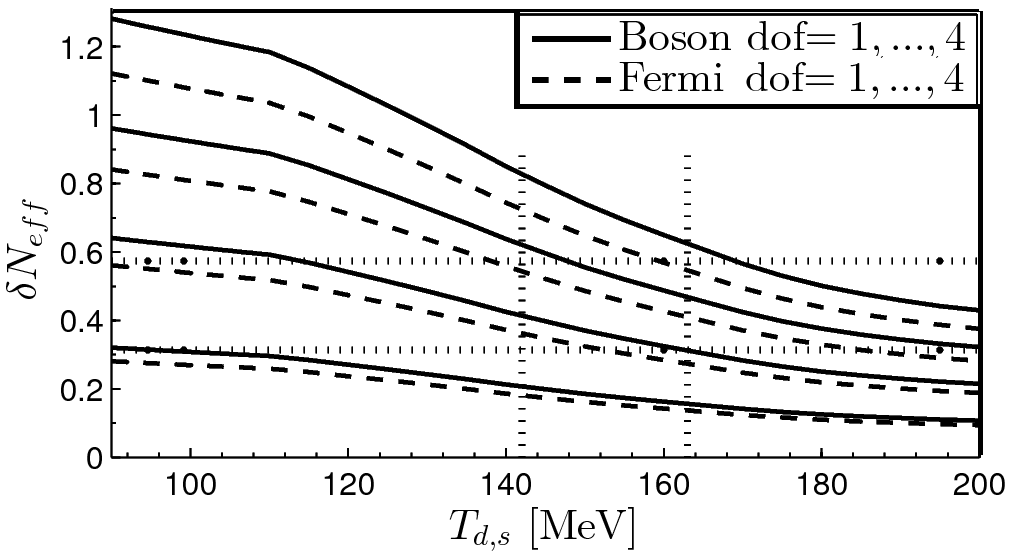}
\caption{Solid lines bottom to top: Increase in $\delta\Neff$ as a function of decoupling point $T_{d,s}$ due to the effect of $1,\dots,4$ light secret boson degrees of freedom ($g_s=1,\dots,4$), and dashed lines, the same for light fermion DoF $(g_s=7/8\times 1,\dots,$ $7/8\times 4)$. The horizontal dotted lines correspond to $\delta\Neff+0.046$=0.36, 0.62  . The vertical dotted lines show the reported range of QGP transformation temperatures ${\Th}\in(142,163)\MeV$.}
\label{fig-delNeff}  
\end{figure}

There is a constraint from above as well, which could be a problem for Ref.\,\cite{Weinberg:2013kea}: if the decoupling occurs at a temperature that is too high, the contribution a particle would make to $\delta N_\mathrm{eff}$ could be too small to matter. We believe that the QGP hadronization point \Th\ is the only special characteristic point and thus this is what we consider in the following as a source condition from where a new secret particle would free-stream. Using the best lattice results to describe the domain of phase transformation of QGP we obtained~\cite{Birrell:2014cja} the contribution of unseen degrees of freedom to $ \delta N_\mathrm{eff}$ for both fermions and bosons as shown in figure~\ref{fig-delNeff}, as a function of the decoupling temperature $T_{d,s}$, where solid lines are for bosons. A smaller (dashed) contribution of fermions is due to their $g_s=7/8\times$ smaller impact. Note that a sterile neutrino would count as two fermionic degrees of freedom.

The horizontal dotted lines in figure~\ref{fig-delNeff} show the domain $\delta\Neff+0.046\in (0.36,0.62) $ compatible with CMB data analysis.  A contribution to the total $\delta\Neff$ of  $0.046$ is expected by mechanisms we will discuss below, and thus the horizontal lines indicate the remaining contribution that the secret particles would need to make. The vertical dotted lines bracket the range of QGP transformation temperatures $ {\Th}\in (142,163)\MeV$. We see that decoupling of 2-3 new particle degrees of freedom could improve the agreement between CMB data and theory of cosmological evolution. We also note that, as the decoupling temperature of the QGP increases, more secret degrees of freedom are needed to explain the presumed CMB signal.  This is  due to the relatively large reheating that occurs after decoupling, as we described earlier. Another  feature is that for decoupling at $T=75\MeV$ we still see significant dillution due to ongoing  reheating by pions and muons.

The possibility to explain data does not mean that there are secret particles that couple to (anti)quarks at hadronization of QGP, even if this hypothesis gives this consideration a sense of plausibility. However, there are already a few eternally searched for particles: axions, and, sterile neutrinos, so first we must clarify if these particles could do the job. Of the two, axions were introduced to protect parity symmetry of strong interactions and thus many firmly believe in their existence. In addition these particles are naturally related to strong interactions. To build a connection to QGP, consider if axions could couple to color field lines, which do not exist in stationary baryons or mesons. Color field lines are abundant in the deconfined QGP state. Therefore, axions could naturally decouple at \Th, which would fit the above picture. And, this would have immediate laboratory consequence, with any QGP fireball possibly shining brightly;  even if the axion ``light" is invisible, this could be noticed due to missing energy~\cite{Birrell:2014cja} if these particles have time to chemically equilibrate in the QGP phase.

It can be safely said that strong interactions have plenty of unexplained challenges and hence the possibility of secret particle freeze-out at \Th\ is a viable hypothesis to be pursued in laboratory experiments searching for missing energy. The CERN experiment SHINE (NA61) -- which in its name already shines secret radiation, has informally expressed interest in pursuing this opportunity. On the other end, and less laboratory verifiable, are speculations about novel symmetry breaking mechanisms specific to the early Universe. It is best here to quote Steven Weinberg\rq s abstract~\cite{Weinberg:2013kea}: \lq\lq It is suggested that Goldstone bosons may be masquerading as fractional cosmic neutrinos, contributing about 0.39 to what is reported as the effective number of neutrino types in the era before recombination. The broken symmetry associated with these Goldstone bosons is further speculated to be the conservation of the particles of dark matter.\rq\rq

\subsection{Degrees of freedom and neutrino freeze-out}\label{nufreeze}
The well-established effect, known to make a contribution $\delta\Neff=0.046$, is that the electron-positron annihilation sends entropy, albeit in a small amount, into the neutrino degrees of freedom. In fact, in figure~\ref{fig-qTnuclear} the domain of neutrino freeze-out is indicated by a vertical line and it is clear that some annihilation is already underway before neutrinos decouple and one must remember that the decoupling process is gradual and energy dependent. We have studied this effect using a novel numerical method~\cite{Birrell:2014gea}.

Since there is a small $\delta\Neff$ effect, one must ask if, and under which conditions, a larger effect could arise -- which would require that in some way our extrapolation back in time to the neutrino decoupling is not entirely correct, perhaps because the conditions are so much more extreme. In fact, if we could turn up alone the mass of the electron without a change in other dimensioned natural constants that could perhaps do the trick nicely and easily as we see inspecting figure~\ref{fig-qTnuclear}: an electron of several MeV would   undergo pair annihilation before the computed neutrino decoupling. However, the computation of neutrino decoupling should change as well since now there are fewer electrons and positrons to scatter from. So the evaluation of what will happen when physical constants change, even one out of many, is not a trivial enterprise.

\begin{figure}
\centering
\sidecaption
\includegraphics[width=0.7\columnwidth,clip]{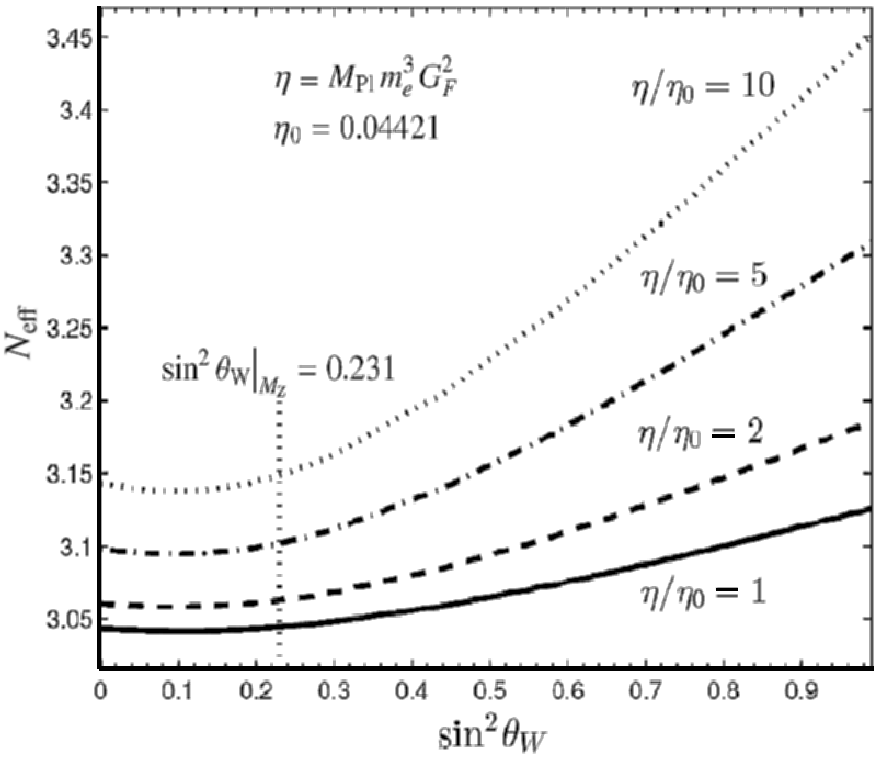}
\caption{The dependence of \Neff\ on the combined parameters of the neutrino freeze-out model in the early Universe, as a function of $\sin^2 \theta_\mathrm{W}$, with $\eta=1,2,5,10$.}
\label{fig-NeffParam}
\end{figure}

We actually needed to perform a dimensional reduction which revealed that there are two parameters that count~\cite{Birrell:2014uka}. The first one is the dimensionless Weinberg angle of electroweak interactions, which controls detail of neutrino cross sections, with the known value obtained from the mass ratio
\begin{equation}\label{eq:sinW}\displaystyle
\frac{M_W}{M_Z} \equiv \cos\theta_\mathrm{W} = 0.8815\;,\ \longrightarrow\ 
\left.\sin^2 \theta_\mathrm{W}\right|_{M_Z}=0.231\;.
\end{equation}
There is no anchor in the Standard Model of particle physics fixing $\sin^2 \theta_\mathrm{W}$ and yet it is difficult to imagine that the ratio of masses in \req{eq:sinW} could be modified in the early Universe. However, the value of $\sin^2 \theta_\mathrm{W}$ varies with the running energy scale relatively strongly and thus $\sin^2 \theta_\mathrm{W}$ could be sensitive to the temperature of the hot Universe at the low scales at which we are studying the decoupling process, without altering the elementary mass ratio that is governed by the $10^5$ times higher energy scale. Hence we show the key results in figure~\ref{fig-NeffParam} as a function of $\sin^2 \theta_\mathrm{W}$.

The other parameter on which the freeze-out condition depends combines all dimensioned variables as follows
\begin{equation}\label{eq:etaMp}
\eta\equiv M_\mathrm{Pl}\, m_e^3\, G_F^2\;, \qquad M_\mathrm{Pl}^2\equiv \frac{1}{8\pi G_N}=2.4354\,10^{18}\,{\rm GeV}\;, 
\quad \eta_0 = 0.04421\;.
\end{equation}
and we find that as $\eta $ grows, the temperature at which the neutrino decoupling occurs drops and thus more of the annihilation entropy of electrons and positrons finds its way into neutrinos, hence increasing $\Neff$. This is so since we can think that increasing $\eta $ the Fermi coupling increases and this makes neutrino scattering on electrons stronger. It helps also that this increase could mean that the mass of the electron is increasing and thus the electron mass moves right into the neutrino decoupling scale while the increase of $G_F$ moves the decoupling towards the electron mass. Similarly, an increase in Planck mass decreases the strength of gravity, and makes the speed of expansion slower. The Planck mass provides the scale of time in which the processes that we are looking at occur; the characteristic time constant is a small fraction of $\tau\equiv M_\mathrm{Pl}/m_e^2=6.18$\,s. Increasing $M_\mathrm{Pl}$ we thus give the system more time to relax and this helps to move entropy into neutrinos. All these remarks are summarized in figure~\ref{fig-NeffParam} which presents the change in \Neff\ for the value $\eta/\eta_0=1,2,5,10$ as a function of $\sin^2 \theta_\mathrm{W}$.

The result shows that it is rather difficult to increase sufficiently the value of \Neff. The difficulty of the situation is that in principle we should believe that there are only two scales, Planck $M_\mathrm{Pl}$ and  Higgs $v$, and the coefficient  $\eta\propto M_\mathrm{Pl}/v$ scales linearly with the ratio of both scales. It is hard to imagine that the Higgs scale which defines the vacuum state can be drastically modified. Similarly, it is difficult to see how there can be a change in the Planck scale allowing  accommodation of a larger value of  \Neff, without other early Universe effects,  for example, a modfied dynamics of nucleosynthesis during the immediately following BBN era. Thus, remembering that BBN works, our finding is that  an allowable change in natural constants could not explain an increase in \Neff. 

As always, there is an \lq if\rq. What if the mass of the electron is not really in the realm of Higgs minimal coupling? If not, it is less fixed  by far and could easily change independently of other constants of nature. This would even help explain why BBN needs not worry about electron density, which as we saw in figure~\ref{fig-qTnuclear}, is significant during the BBN time period. So if for example $m_e$ could be three times greater and of the scale of the neutrino decoupling temperature, that would make the situation rather different for \Neff. A reason to think in this way is that the $m_e$ scale looks very \lq electromagnetic\rq: take any electromagnetic mass difference for any of particles around and one sees that 1.5 MeV is a typical variation. Also note that the required minimal coupling of Higgs to electron is exceptionally weak, $g=0.5/250 000=2\,10^{-6}$. The experimental limit on  Higgs-electron coupling is 600 times above that expected  by minimal coupling~\cite{Altmannshofer:2015qra}; thus we do not really know if electron mass originates in the Higgs mechanism. Moreover, the expected small value of $g$  renders higher order effects involving  other interactions relevant, an example of this type of behavior is the fact that in an electron-positron collider Higgs would not be produced by minimal interaction, but by another  higher order coupling to Higgs, see for example Ref.\,\cite{Han:2015ofa}.

The conclusion of this subsection is that only if the electron is not really what Higgsologists want it to be, then there is a chance that we can, in agreement with all other early Universe considerations, create a scenario such that \Neff\ is modified by an increase in the electron mass present in the distant past at high temperature. 

\section{Conclusions}
Hagedorn and his temperature were of pivotal importance in the development of the modern picture of the early Universe. Today, the experimental and theoretical insights into the working of the early Universe and the connection of QGP and Hagedorn hadron phase offer an interesting opportunity, a fulfillment of the promise that relativistic heavy ion collisions are going to shine new light on the early Universe physics. 

In the near future, experimental precision of the CMB data should improve greatly. Thus it makes good sense to characterize the evolution of the Universe from the present through to the QGP era under the most conservative assumptions about its properties, so as to create the model background against which the new data can be tested. An interesting outcome of this exercise we have obtained is an overview of the composition of the Universe across all evolution epochs and have shown here how the entropic degrees of freedom evolve. 

One of the most striking outcomes is the strong dependence of the available degrees of freedom on the QCD-QGP equations of state. We believe that we are first to employ realistic lattice-QCD results in this domain and were able to present here a glimpse on our extensive study of the evolution of the Universe connecting the QGP era to the observational period after recombination. 

The connection to observables involves the passage through the neutrino decoupling era, requiring a refined understanding of neutrino freeze-out, electron positron annihilation and the dependence of these processes on natural constants. Within the realm of conventional physics we have been able to solve this problem in full. We conclude that we can use these results to study the QGP phase of the early Universe, and in particular to explore it it can be a natural source of undiscovered light particles.

The interesting worry impacting this plan of action is, whether the electron is just another minimally coupled (to Higgs) elementary particle, or if its small mass signals some novel physics. If electron mass is not constant as a function of time and is also sensitive to  temperature in the context of the early Universe, then our conclusions will be modified. On the other hand we gain insight into another related physics domain, the unraveling of the electron,  and this is very much worth the effort.

\appendix
\section{The ${\uLambda}$CDM standard model}\label{cosmo}
\subsection{Standard Cosmology}\label{scosmo}
In order to travel back in time to the period when QGP dominated the Universe and beyond, we first need to elaborate on the relation between the expansion dynamics of the Universe and temperature. For this purpose we need some preparation in the standard cosmological (FLRW-Universe) model. We use the spacetime metric
\beqn\label{metric}
ds^2=c^2dt^2-a^2(t)\left[ \frac{dr^2}{1-kr^2}+r^2(d\theta^2+\sin^2(\theta)d\phi^2)\right]
\eeqn
characterized by the scale parameter $a(t)$ of a spatially homogeneous Universe. The geometric parameter $k$ identifies the geometry of the spacial hypersurfaces defined by comoving observers. Space is a flat-sheet for the observationally preferred value $k=0$ \cite{Ade:2013zuv}. In this case it can be more convenient to write the metric in rectangular coordinates
\beqn\label{metric2}
ds^2=c^2dt^2-a^2(t)\left[ dx^2+dy^2+dz^2\right].
\eeqn
We will work in units where $\hbar=1,c=1$. 

The global Universe dynamics can be characterized by two quantities: the Hubble parameter $H$, a strongly time dependent quantity on cosmological time scales, and the deceleration parameter $q$:
\beqn\label{dynamic}
\frac{\dot a }{a}\equiv H(t) ,\quad \frac{\ddot a}{a}=-qH^2,\quad 
q\equiv -\frac{a\ddot a}{\dot a^2},\quad \dot H=-H^2(1+q). 
\eeqn
The Einstein equations are: 
\beqn\label{Einstine}
G^{\mu\nu}=R^{\mu\nu}-\left(\frac R 2 +{\uLambda}\right) g^{\mu\nu}=8\pi G_N T^{\mu\nu}, 
\quad R= g_{\mu\nu}R^{\mu\nu}.
\eeqn
Symmetry considerations imply that the stress energy tensor is determined by an energy density and an isotropic pressure
\begin{align}
 T^\mu_\nu =\mathrm{diag}(\varepsilon, -P, -P, -P).
\end{align}
It is common to absorb the Einstein cosmological constant ${\uLambda}$ into the energy and pressure
\beqn\label{EpsLam}
\varepsilon_{\uLambda}=\frac{{\uLambda}}{8\pi G_N}=25.6\meV^4, \qquad P_{\uLambda}=-\varepsilon_{\uLambda}
\eeqn
and we implicitly consider this done from now on. The fitted value of the dark energy is indicated 

One should note that the bag constant $ {\cal B} $ of the quark-bag model has the same behavior in regard to energy and momentum as has the Einstein cosmological parameter $ {\cal B} \leftrightarrow {\uLambda}/8\pi G_N$. $ {\cal B} $ adds positively to the energy density but negatively to the pressure, counteracting the positive particle pressure. Contrary to the initial expectation based on the quark-bag model where quark pressure is in equilibrium with the bag constant, in the dynamical Universe the appearance of such a bag term will accelerate the expansion just like today where we see an acceleration due to dark energy. The parallel meaning of $ {\cal B}$ and $ {\uLambda}/8\pi G_N$ relies on both quantities acting within the volume of their respective `Universe', in the sense that $ {\cal B}$ is strictly and only present within the volume of quark blobs -- hadrons, or QGP.

Two dynamically independent equations arise using the metric \req{metric} in \req{Einstine}:
\beqn\label{hubble}
\frac{8\pi G_N}{3} \varepsilon = \frac{\dot a^2+k}{a^2}
=H^2\left( 1+\frac { k }{\dot a^2}\right),
\qquad
\frac{4\pi G_N}{3} (\varepsilon+3P) =-\frac{\ddot a}{a}=qH^2.
\eeqn
We can eliminate the strength of the interaction, $G_N$, solving both these equations for ${8\pi G_N}/{3}$, and equating the result to find a relatively simple constraint for the deceleration parameter:
\beqn\label{qparam}
q=\frac 1 2 \left(1+3\frac{P}{\varepsilon}\right)\left(1+\frac{k}{\dot a^2}\right).
\eeqn
For a spatially flat Universe, $k=0$, note that in a matter-dominated era where $P/\varepsilon<<1$ we have $q\simeq 1/2$; for a radiative Universe where $3P=\varepsilon$ we find $q\simeq 1 $; and in a dark energy Universe in which $P=-\varepsilon$ we find $q=-1$. Spatial flatness is equivalent to the assertion that the energy density of the Universe equals the critical density
\begin{equation}\label{crit_density}
\varepsilon=\varepsilon_{\text{crit}}\equiv \frac{3H^2}{8\pi G_N}.
\end{equation}

The CMB power spectrum is sensitive to the deceleration parameter and the presence of spatial curvature modifies $q$. The Planck results~\cite{Ade:2013zuv} constrain the effective curvature energy density fraction,
\begin{equation}
\Omega_K\equiv1-\varepsilon/\varepsilon_{\text{crit}},
\end{equation} 
to
\begin{equation}
|\Omega_K|<0.005.
\end{equation}
This indicates a nearly flat Universe. We will work here within an exactly spatially flat cosmological model, $k=0$.

As must be the case for any solution of Einstein's equations, \req{hubble} implies that the energy momentum tensor of matter is divergence free:
\beqn\label{divTmn}
T^{\mu\nu};_\nu =0 \Rightarrow -\frac{\dot\varepsilon}{\varepsilon+P}=3\frac{\dot a}{a}=3H.
\eeqn
A dynamical evolution equation for $\varepsilon(t)$ arises once we combine \req{divTmn} with \req{hubble}, eliminating $H$. Given an equation of state $P(\varepsilon)$, solutions of this equation describes the dynamical evolution of matter in the Universe. In practice, we evolve the system in both directions in time. On one side, we start in the present era with the energy density fractions fit by Planck~\cite{Ade:2013zuv},
\begin{equation}\label{Planck_params}
H_0=67.74\text{km/s/Mpc}\;,\hspace{2mm} \Omega_b=0.05\;,\hspace{2mm} \Omega_c=0.26\;, \hspace{2mm}\Omega_{\uLambda}=0.69\;,
\end{equation}
and integrate backward in time. On the other hand, we start in the QGP era with an equation of state determined by an ideal gas of SM particles, combined with a perturbative QCD equation of state for quarks and gluons, and integrate forward in time.

\subsection{Matter Content}\label{mcosmo}
In this work, matter will be modeled by a particle distribution function $f(t,x,p)$ that, roughly speaking, gives the probability of finding a particle per unit spacial volume per unit momentum space volume at a given time. The distribution function gives the stress energy tensor, particle four-current, and entropy four-current via 
\begin{align}
T^{\mu\nu}(t,x)=&\frac{d}{(2\pi)^3}\int p^\mu p^\nu f(t,x,p) \sqrt{|g|}\frac{d^3p}{p_0},\\
n^\nu(t,x)=&\frac{d}{(2\pi)^3}\int p^\nu f(t,x,p) \sqrt{|g|}\frac{d^3p}{p_0},\\
s^\nu(t,x)=&-\frac{d}{(2\pi)^3}\int(f\ln(f)\pm(1\mp f)\ln(1\mp f))p^\nu\sqrt{|g|}\frac{d^3p}{p_0}\;,
\end{align}
where the upper signs are for fermions, the lower for bosons, $d$ is the degeneracy of the particle, and $g$ is the determinant of the metric. In a flat FRW Universe, the expressions for the energy density, pressure, number density, and entropy density of a particle of mass $m$ are
\begin{align}\label{moments}
\varepsilon=&\frac{d}{(2\pi)^3}\int f(t,x,p)Ed^3p,\\
P=&\frac{d}{(2\pi)^3}\int f(t,x,p)\frac{p^2}{3E}d^3p,\\
n=&\frac{d}{(2\pi)^3}\int f(t,x,p) d^3p, \hspace{2mm} E=\sqrt{m^2+p^2},\\
s=&-\frac{d}{(2\pi)^3}\int (f\ln(f)\pm(1\mp f)\ln(1\mp f)) d^3p.
\end{align}

\subsection{Equilibrium and Free-Streaming Distribution}\label{fcosmo}
Free-streaming is a type of non-equilibrium distribution that is significant in cosmology. Here we outline its properties, including what distinguishes it from the equilibrium distributions. The freeze-out process, whereby a particle species stops interacting and decouples from the photon background, involves several steps that lead to the final form of the free-streaming momentum distribution. For further details, see Ref.\,\cite{Birrell:2012gg}.

Chemical freeze-out of a particle species occurs at the temperature, $T_\mathrm{ch}$, when particle number changing processes slow down and the particle abundance can no longer be maintained at an equilibrium level. Prior to the chemical freeze-out temperature, number changing processes are significant and keep the particle in chemical (and thermal) equilibrium, implying that the distribution function has the Fermi-Dirac form, obtained by maximizing entropy at fixed energy
\begin{equation}\label{equilibrium}
f_{c}(t,E)=\frac{1}{\exp(E/T)+1}, \text{ for } T(t)> T_\mathrm{ch}.
\end{equation}

Kinetic freeze-out occurs at the temperature, $T_f$, when momentum exchanging interactions no longer occur rapidly enough to maintain an equilibrium momentum distribution. When $T_f<T(t)<T_\mathrm{ch}$, number changing process no longer occur rapidly enough to keep the distribution in chemical equilibrium but there is still sufficient momentum exchange to keep the distribution in thermal equilibrium. The distribution function is therefore obtained by maximizing entropy, with a fixed energy, particle number, and antiparticle number separately, implying that the distribution function has the form
\begin{equation}\label{kinetic_equilib}
f_k(t,E)=\frac{1}{\Upsilon^{-1}\exp(E/T)+1}, \text{ for }T_f< T(t)< T_\mathrm{ch}.
\end{equation}
The fugacity
\begin{equation}
\Upsilon(t)\equiv e^{\sigma(t)}
\end{equation}
controls the occupancy of phase space and is necessary once $T(t)<T_\mathrm{ch}$ in order to conserve particle number. See Ref.\,\cite{Birrell:2012gg} for a detailed discussion of its significance.

For $T(t)<T_f$ there are no longer any significant interactions that couple the particle species of interest and so they begin to free-stream through the Universe, i.e. travel on geodesics without scattering. The Einstein-Vlasov equation can be solved, see Ref.\,\cite{choquet2008general}, to yield the free-streaming momentum distribution
\begin{equation}\label{free_stream_dist}
f(t,E)=\frac{1}{\Upsilon^{-1}e^{\sqrt{p^2/T^2+m^2 /T_f^2}}+ 1}
\end{equation}
where the free-streaming effective temperature
\begin{equation}\label{T_freestream_dist}
T(t)=\frac{T_fa(t_k)}{a(t)}
\end{equation}
is obtained by red-shifting the temperature at kinetic freeze-out.

The corresponding free-streaming energy density, pressure, and number densities are given by
\begin{align}
\varepsilon&=\frac{d}{2\pi^2}\!\int_0^\infty\!\!\!\frac{\left(m^2+p^2\right)^{1/2}p^2dp }{\Upsilon^{-1}e^{\sqrt{p^2/T^2+m^2/T_f^2}}+ 1},\label{freestream_rho}\\[0.2cm]
P&=\frac{d}{6\pi^2}\!\int_0^\infty\!\!\!\frac{\left(m^2+p^2\right)^{-1/2}p^4dp }{\Upsilon^{-1} e^{\sqrt{p^2/T^2+m^2/T_f^2}}+ 1},\label{freestream_P}\\[0.2cm]
n&=\frac{d}{2\pi^2}\!\int_0^\infty\!\!\!\frac{p^2dp }{\Upsilon^{-1}e^{\sqrt{p^2/T^2+m^2/T_f^2}}+ 1},
\label{num_density}
\end{align}
where $d$ is the degeneracy of the particle species. These differ from the corresponding expressions for an equilibrium distribution in Minkowski space by the replacement $m\rightarrow m T(t)/T_f$ {\em only} in the exponential. 

The separation of the freeze-out process into these three regimes is of course only an approximation. In principle there is a smooth transition between them. However, it is a very useful approximation in cosmology. See \cite{Mangano:2005cc,Birrell:2014gea} for methods capable of resolving these smooth transitions.


\end{document}